\documentclass[aps,preprint]{revtex4}
\usepackage{epsfig}

\def\[{\left[}
\def\]{\right]}
\def\({\left(}		
\def\){\right)}		
\def\bar{\overline}
\def\zbar{{\bar{z} }}
\def\inv#1{{1\over #1}}
\def\d{\partial}
\def\ep{\epsilon}
\def\vep{\varepsilon}
\def\CF{{\cal F}}
\def\CH{{\cal H}}
\def\CO{{\cal O}}
\def\CP{{\cal P}}
\def\CT{{\cal T}}
\def\barray{\begin{eqnarray}}
\def\earray{\end{eqnarray}}
\def\beq{\begin{equation}}
\def\eeq{\end{equation}}
\def\chidag{\chi^\dagger}
\def\dim#1{\lbrack\!\lbrack #1 \rbrack\!\rbrack }
\def\chidagger{\chi^\dagger}
\def\xvec{{\bf x}}
\def\gammachi{{\gamma_\chi}}
\def\gammam{{\gamma_m}}
\def\absx{|\xvec|}
\def\kvec{{\bf k}}
\def\nvec{{\vec{n}}}
\def\chibar{{\bar{\chi}}}
\def\daggerc{{\dagger_c}}
\def\intk{\frac{ d^d \kvec }{(2\pi)^d }}
\def\Tr{{\rm Tr}}

\begin{document}

\preprint{MZ-TH/07-09}

\title{Semi-Lorentz invariance, unitarity, and critical exponents of symplectic fermion models}

\author{Andr\'e LeClair$^a$ and Matthias Neubert$^{a,b}$}

\affiliation{$^a$\,Laboratory of Elementary-Particle Physics, Cornell University, Ithaca, New York 14853, U.S.A.\\
$^b$\,Institut f\"ur Physik (THEP), Johannes Gutenberg-Universit\"at,
D--55099 Mainz, Germany}

\date{May 30, 2007}

\begin{abstract}
We study a model of $N$-component complex fermions 
with a kinetic term that is second order in derivatives.
This symplectic fermion model has an $Sp(2N)$ symmetry, which for 
any $N$ contains an $SO(3)$ subgroup that 
can be identified with rotational spin of 
spin-$\inv{2}$ particles.   Since the spin-$\inv{2}$ representation
is not promoted to a representation of the Lorentz group,
the model is not fully Lorentz invariant, although it has
a relativistic dispersion relation.    The hamiltonian
is pseudo-hermitian, $H^\dagger = C H C$,  which implies it
has a unitary time evolution.   Renormalization-group analysis 
shows the model has a low-energy fixed point that is 
a fermionic version of the Wilson-Fisher fixed points. The
critical exponents are computed to two-loop order.  
Possible applications to condensed matter physics in 
$3$ space-time dimensions are discussed.   
\end{abstract}

\maketitle

\section{Introduction}

A basic result in the quantum field theory of
fundamental particles in $4$ space-time dimensions is that requiring  Lorentz
invariance for spin-$\frac12$ particles necessarily
leads to the free  Dirac lagrangian.  Here ``spin-$\frac12$''
refers to the 3-dimensional rotational subgroup of
the Lorentz group.   The Lie algebra of the Lorentz
group is $SU(2)  \otimes SU(2)$, where one linear combination 
of the two $SU(2)$ symmetries is identified as angular momentum; 
thus spin representations are  promoted to Lorentz representations
in a straightforward manner.  

This paper in part deals with the following basic question. 
Suppose one wishes to describe the quantum field theory of
spin-$\frac12$ particles in a physical context where the 
dispersion relation happens to be Lorentz invariant, but unlike
in fundamental particle physics
the full Lorentz invariance is not evidently required.    
In particular, we have
in mind systems in condensed matter physics where the
relativistic dispersion relation arises as a consequence of
special properties of the system,  and what plays the role of
the speed of light is some material-dependent velocity.  For instance,
the effective mass of electronic quasi-particles may go zero
because one is near a quantum critical point or more
simply because of band structure as in 2-dimensional graphene 
\cite{Geim,Kim},
and massless particles, like photons,  
usually require a relativistic
dispersion relation.   Under these circumstances, the question is
whether such a quantum field theory must necessarily be that of a
Dirac fermion.    
In the case of graphene,  the fermionic
quasi-particles do turn out to be described by the massless Dirac 
equation. The reason for this is not an intrinsic Lorentz
invariance, but rather that the continuum limit of a
tight binding model on a hexagonal lattice gives a hamiltonian
that is first order in derivatives, and near the Fermi points
the particles are massless.    

In this paper we will study an alternative
to the Dirac theory.  
The model  is built out of  an $N$-component 
complex fermionic field with
a non-Dirac, two-derivative action with a Lorentz-invariant  
dispersion relation.  This model has a symplectic $Sp(2N)$ symmetry. 
If this symmetry is viewed as an internal symmetry, then the
fields are Lorentz scalars and the theory is Lorentz invariant
but with the ``wrong'' statistics.  However,     
since the Lie group $Sp(2N)$ has an $SO(3)$ subgroup,  we can identify
the latter with rotational spin.    Therefore our model can naturally
describe spin-$\frac12$ particles.  The fermionic statistics 
is then in  accordance with the spin-statistics
connection, which requires spin-$\frac12$ particles to be described by
fermionic fields.  However since the rotational spin-$\frac12$ is not promoted
to a representation of the Lorentz group as in the Dirac theory, our 
model is 
strictly speaking not Lorentz invariant;  hence the terminology
``semi-Lorentz'' invariant.  

The most serious potential problem of this theory concerns its
unitarity,  and this is addressed in the present paper.    We show that
the hamiltonian is pseudo-hermitian,
\beq\label{intro.1}
   H^\dagger = C H C \,,
\eeq
where $C$ is a unitary operator satisfying $C^2=1$.  This
generalization of hermiticity was considered long ago by
Pauli \cite{Pauli},  and more recently by Mostafazadeh 
\cite{Mostafazadeh}
as a way of explaining the real spectrum and addressing 
the unitarity issue  in $\CP\CT$ symmetric 
quantum mechanics \cite{Bender1,Bender2}.   The important point is that
by suitably defining a $C$-dependent inner product,  pseudo-hermiticity
of $H$ is sufficient to ensure a unitary (i.e., norm-preserving) 
time evolution. This is explained in section~II.  
One  should also point out that taking a non-relativistic limit in
the kinetic term one obtains a perfectly unitary second-quantized
description of interacting fermions. 

The identification of the $SO(3)$ Lie sub-algebra of $Sp(2N)$ with
rotational spin is described in section~III.   There we also
study the discrete space-time symmetries of time-reversal and parity,
and show how the spin generators transform properly under them. 

Another possible signature of non-unitarity comes from studying
finite-size or finite-temperature effects. 
We show in section~IV that whereas
imposing periodic boundary conditions leads to a negative coefficient
in the free energy,  correctly imposing anti-periodic boundary 
conditions,  as is normally appropriate for fermions,  leads to 
a positive coefficient.  (In two dimensions this coefficient 
is related to the Virasoro central charge.) 

Symplectic fermions are interesting for potential applications
to critical phenomena.   First of all, in $D=3$,  
since the group of spacial rotations is simply $U(1)$, 
there are less constraints coming from Lorentz invariance.  
More importantly, unlike Dirac fermions,   
four-fermion interactions in $D=3$ drive the theory to some
novel low-energy fixed points that are fermionic versions
of the Wilson-Fisher \cite{WilsonFisher} fixed points.  
The reason is simple: in three dimensions a symplectic fermion 
$\chi$ has classical (or mean field) 
scaling dimension $1/2$,  whereas Dirac fermions $\psi$ have 
dimension 1;
therefore $\chi^4$ is a dimension-2 operator and is relevant in 
$D=3$, whereas $\psi^4$ is irrelevant.    This was the original 
motivation for the work \cite{spinon}, where it was attempted to  interpret
these fixed points at $N=2$  as  examples of deconfined quantum 
criticality \cite{Senthil}.     Although it remains unclear 
whether our fermionic critical point can correctly describe  deconfined
quantum criticality as defined in \cite{Senthil},  the
resolution of the unitarity problem as presented in this paper 
is  sufficient motivation to analyze the critical exponents further. 
In sections V and VI we extend the analysis of \cite{spinon} 
to two-loop order.
In particular, we show that some of the critical exponents can be
obtained by analytically continuing known results for the
$O(M)$ Wilson-Fisher fixed point to $M=-2N$ \footnote{This feature
was not properly appreciated in \cite{spinon} due to 
an error by a factor of 2 in the critical exponent $\gamma_\chi$.}. 
We also calculate the critical exponents for composite bilinear
operators, which to our knowledge have not been studied for
the $O(M)$ fixed points \footnote{The anomalous dimension 
of such bilinear operators was incorrectly assumed, to lowest order, 
to be twice
the dimension of the fundamental field $\chi$ in \cite{spinon}.}.

The correspondence of our model with $O(M)$ for negative $M$ is 
merely formal and not
expected to be valid for all physical properties.  First of all,
the symmetries of the models are different.      
It should also be clear from the fact that in applications to
condensed matter physics, our model has a Fermi surface, etc.  
Some concrete distinctions in the finite-size effects are 
made in section~IV. 

In section~VII we speculate on some possible applications 
to $2+1$ dimensional quantum criticality in condensed matter
physics.   In the broadest terms,  at the fixed point our model can be viewed
as a quantum critical theory of spinons,  where the symplectic
fermions  are fundamental spinon fields.   For $N=2$ components, 
we discuss 
possible applications to quantum anti-ferromagnetism, where the 
magnetic order parameter $\nvec$ is a composite operator 
in terms of spinons $\nvec=\chidagger\vec{\sigma}\,\chi$.
This compositeness  is the same as in  deconfined quantum 
criticality \cite{Senthil},  however our model is different
since it has no $U(1)$ gauge field.   We show that two-point
correlation exponents ($\eta$) are rather large compared to
the bosonic Wilson-Fisher fixed point, and this is mainly due
to the compositeness of $\nvec$.   By treating both cases,  
we show this  is true irrespective
of whether the particles are bosons or fermions.  

Section~VIII contains a summary of our main findings and some conclusions.

\section{Pseudo-hermiticity and unitarity  of complex scalar fermions}

Let $\chi$ denote an $N$-component complex 
field and consider the following action in $D=d+1$ dimensional 
Minkowski space:
\beq\label{sp1}
   S_\chi 
   = \int d^d\xvec\,dt \left[
   \d^\mu\chibar \d_\mu\chi - m^2 \,  \chibar\chi
   - 4\pi^2 g\,(\chibar \chi)^2 \right] ,
\eeq
where $\chibar\chi=\sum_{\alpha=1}^N \chibar^\alpha\chi^\alpha$,
and $\d^\mu\d_\mu=\d_t^2-\d_\xvec^2$.  
If $\chi$ is taken to be a  Lorentz scalar, then the model 
is Lorentz invariant.    
The above action has an explicit internal $U(N)$ symmetry. 
In the next section we show that, if $\chi$ is a fermion,
then there is actually a hidden $Sp(2N)$ symmetry.     

We wish to quantize this model with $\chi$ taken to
be a fermionic (Grassman) field.    The conjugate-momentum
fields are $\pi_\chi=-\d_t\chibar$ and
$\pi_\chibar=\d_t\chi$. They obey the canonical
anti-commutation relations
\beq\label{tw1}
   \{ \chibar^\alpha(\xvec,t), \d_t\chi^\beta(\xvec',t) \} 
   = - \{ \chi^\alpha(\xvec,t), \d_t\chibar^\beta(\xvec',t) \} 
   = i\delta^{\alpha\beta} \delta^{(d)}(\xvec-\xvec') \,.
\eeq
The hamiltonian for this system is 
\beq\label{tw2}
   H = \int d^d\xvec \left[
   \d_t\chibar \d_t\chi + \d_\xvec\chibar \d_\xvec\chi 
   + m^2 \, \chibar\chi + 4\pi^2 g \,  (\chibar \chi)^2 \right] .
\eeq
Note that because of the
fermion statistics the interaction term vanishes for $N=1$.

Consider first the free theory with $g=0$.  
Suppressing the component indices, the fields have the following 
mode expansions consistent with the equations of motion:
\barray\label{tw3}
   \chi(x) 
   &=& \int\frac{d^d\kvec}{(2\pi)^{d/2}\sqrt{2\omega_\kvec}}
    \( b_{\kvec,-}^\dagger\,e^{-ik\cdot x}
     + b_{\kvec,+}\,e^{ik\cdot x} \) , \nonumber\\ 
   \chibar(x) 
   &=& \int\frac{d^d\kvec}{(2\pi)^{d/2}\sqrt{2 \omega_\kvec}}
    \( - b_{\kvec,-}\,e^{ik\cdot x}
     + b_{\kvec,+}^\dagger\,e^{-ik\cdot x} \) , 
\earray
where $\omega_\kvec = \sqrt{\kvec^2 + m^2 }$ and 
$k\cdot x = \omega_\kvec\,t - \kvec \cdot \xvec$.
The extra minus sign in the $b_-$ term in $\chibar$ is chosen so that
the anti-commutation relations (\ref{tw1}) lead to the standard non-vanishing canonical relations
\beq\label{tw4}
   \{ b_{\kvec+}, b^\dagger_{\kvec'+} \}
   = \{ b_{\kvec-}, b^\dagger_{\kvec'-} \} 
   = \delta^{(d)} (\kvec - \kvec' ) \,,
\eeq
where it is understood that the field operators belong to the same field component. Of course, the extra sign would be 
unnecessary if $\chi$ was a bosonic field.  

With the above definitions the fields have the required properties
under causality (see for instance \cite{Weinberg}), namely
$\{\chi^\alpha(x), \chi^\beta(y)\}
= \{\chibar^\alpha(x), \chibar^\beta(y)\} = 0$ and
\beq\label{tw4bb}
   \{ \chi^\alpha(x), \chibar^\beta(y) \} 
   = \delta^{\alpha\beta} \left[
   \Delta_+ (x-y) - \Delta_+ (y-x) \right] ,
\eeq
where
\beq\label{tw4bbb}
   \Delta_+(x) = \int\frac{d^d\kvec}{(2\pi)^d\,2\omega_\kvec}\,
   e^{i\kvec\cdot\xvec} \,. 
\eeq
Since $\Delta_+ (x)$ depends only on $x^2$ and is well-defined for
space-like separation ($x^2<0$), it follows that $\chi(x)$ and 
$\chibar(y)$ anti-commute
for space-like separated points  $x$ and $y$.   This implies that the
hamiltonian densities  $\CH(x)$ and $\CH(y)$ also commute at 
space-like separation.    We will return to the spin-statistics connection below.  

In terms of the  momentum-space modes, the free hamiltonian is
\beq\label{tw4b}
   H = \int d^d\kvec ~ \omega_\kvec
   \( b^\dagger_{\kvec+} b_{\kvec+} - b_{\kvec-} b^\dagger_{\kvec-} \)
   = \int d^d\kvec ~ \omega_\kvec 
   \( b^\dagger_{\kvec+} b_{\kvec+} + b^\dagger_{\kvec-} b_{\kvec-}
   \) + \mbox{const.}
\eeq
We can discard the infinite constant in the above equation,
which is equivalent to normal-ordering the hamiltonian. 
Let us define the vacuum $|0\rangle$ as the state being 
annihilated by $b_\pm$.    
As a result,  all states have positive energy.  The one-particle
states are doubly degenerate:
\beq\label{spec}
   H\,b^\dagger_{\kvec,\pm}|0\rangle 
   = H|\kvec,\pm\rangle = \omega_\kvec |\kvec,\pm\rangle \,. 
\eeq

For $N=1$ the theory is manifestly invariant under a $U(1)$ symmetry.  The corresponding conserved current satisfying $\d^\mu J_\mu=0$ is
$J_\mu = i[(\d_\mu\chibar)\chi-\chibar\,\d_\mu\chi]$. 
The conserved charge $Q=\int d^d\xvec\,J_t(\xvec)$ can be expressed 
as 
\beq\label{tw6}
   Q = \int d^d\kvec ~ \omega_\kvec
   \( b^\dagger_{\kvec+} b_{\kvec+} + b_{\kvec-} b^\dagger_{\kvec-}
   \) .
\eeq
With these conventions, the operators 
$b^\dagger_+$ and $b_-$ have charge $Q=1$,
whereas $b^\dagger_-$ and $b_+$ have charge $Q=-1$.  

The usual spin-statistics connection (see for instance 
\cite{Weinberg})
is based on causality as described above, along with the requirement
that the hamiltonian must be constructed out of $\chi$ and
its hermitian adjoint in order for it to be hermitian.  
The latter is violated here:  
because of the extra minus sign in the mode expansion of $\chibar$ 
for the fermionic case,  one sees
that unlike for the bosonic case,  $\chibar$ is not 
the hermitian adjoint of $\chi$.   Let us introduce
a unitary operator $C$ satisfying 
$C^\dagger C =1$ and  $C=C^\dagger$, which is defined by the properties $C b_\pm C=\pm b_\pm$ and 
$C b_\pm^\dagger C=\pm b_\pm^\dagger$. 
Then the relation between $\chibar$ and $\chi$ can be expressed as
\beq\label{tw8} 
   \chibar = C \chi^\dagger C \,.
\eeq
Since $(\chibar\chi)^\dagger = C\chibar\chi C$,  
the hamiltonian satisfies the ``intertwined"  hermiticity condition
\beq\label{tw9}
   H^\dagger = C H C \,.
\eeq
The above is also true in the interacting theory since 
$[(\chibar\chi)^2]^\dagger=C(\chibar \chi)^2 C$.
(Note that for the free theory in momentum space, 
$C$ actually commutes with $H$ as can be seen from eq.~(\ref{tw4b}),
and hence the above equation is trivially satisfied; 
however, this is no longer the case in the interacting theory.)   

Hamiltonians satisfying eq.~(\ref{tw9}) were considered long ago by
Pauli \cite{Pauli} and more recently by Mostafazadeh 
\cite{Mostafazadeh} in connection with $\CP\CT$ symmetric 
quantum mechanics \cite{Bender1,Bender2}.  We will
follow the previously introduced terminology and refer to such
hamiltonians as $C$-pseudo-hermitian.   Quantum mechanics based on 
pseudo-hermitian operators has some very desirable properties,
which parallel the standard ones.   
First of all, as we now explain, a pseudo-hermitian hamiltonian can 
still define a unitary quantum mechanics 
if one defines the inner product appropriately.  Specifically, 
consider a new inner product defined as
\beq\label{tw10}
   \langle\psi'|\psi\rangle_c \equiv \langle\psi'|C|\psi\rangle \,.
\eeq
Then probability is conserved with respect to this modified inner product, i.e., the norms of states are preserved under time 
evolution:
\beq\label{tw11} 
   \langle\psi'(t)|\psi(t)\rangle_c 
   = \langle\psi'| e^{iH^\dagger t} C e^{-iHt} |\psi\rangle 
   = \langle\psi'| C e^{iHt } C^2 e^{-iHt} |\psi\rangle 
   = \langle\psi'|\psi \rangle_c \,.
\eeq
Pseudo-hermiticity also ensures that the eigenvalues of $H$ are real.
To see this, let $|\psi_E\rangle$ denote an eigenstate of $H$ with
eigenvalue $E$. Then
\beq\label{realE}
   (E-E^*)\,\langle\psi_E|\psi_E\rangle_c 
   = \langle\psi_E| (CH-H^\dagger C) |\psi_E\rangle 
   = 0 \,.
\eeq
Therefore eigenstates of $H$ with non-zero C-norm necessarily have
real energies. Other properties are proven in \cite{Mostafazadeh}.  

In the sequel it will be convenient to define the  
pseudo-hermitian adjoint $A^\daggerc$ of any operator $A$ 
as the proper hermitian adjoint with respect to the 
C-inner product:
\beq\label{twistedher}
   \langle\psi'|A|\psi\rangle_c^* 
   \equiv \langle\psi|A^\daggerc|\psi'\rangle_c \,, 
\eeq
which implies
\beq\label{dagc}
   A^\daggerc = C A^\dagger C \,.
\eeq
The pseudo-hermiticity condition on the hamiltonian then simply 
reads $H^\daggerc=H$.
One can easily establish that this pseudo-hermitian adjoint satisfies the usual rules, e.g.\
\barray\label{rules1}
   (AB)^\daggerc &=& B^\daggerc A^\daggerc \,, \nonumber\\
   (a A + b B)^\daggerc &=& a^* A^\daggerc + b^* B^\daggerc \,,
\earray
where $A,B$ are operators and $a,b$ are complex numbers.

\section{$Sp(2N)$ symmetry and rotational spin}

The action (\ref{sp1}) has an explicit $U(N)$ symmetry irrespective 
of whether  
$\chi$ is bosonic or fermionic.   If $\chi$ is bosonic,  
then the action expressed in terms of real fields has an
$O(2N)$ symmetry. On the other hand, if $\chi$ is
a fermionic field, then the symmetry is  
$Sp(2N)$.  Even for $N=1$, the $Sp(2)$ symmetry of the fermionic
theory is larger than the $U(1)$ symmetry of the bosonic theory.
To manifest this symmetry explicitly,
let us express each component $\chi^\alpha$ as
\beq\label{sp2}
   \chi^\alpha
   = \inv{\sqrt{2}}\,(\eta_1^\alpha  + i\eta_2^\alpha) \,, \qquad
   \chibar^\alpha  
   = \frac{-i}{\sqrt{2}}\,(\eta_1^\alpha  - i\eta_2^\alpha) \,. 
\eeq
We now introduce the $2\times 2$ anti-symmetric matrix
$\ep=\Big(\begin{array}{rc} 0~ & 1 \\[-0.2cm] -1~ & 0 
\end{array}\Big)$
and the $2N\times 2N$ matrix $\ep_N=\ep\,\otimes 1_N$.  
Arranging the real fields $\eta_i^\alpha$ into a $2N$ vector
$\eta=(\eta_1^1,\eta_2^1,\eta^2_1,\eta^2_2,\dots,
\eta_1^N,\eta_2^N)$, one has
\beq\label{sp3}
   \chibar\chi = \inv{2}\,\eta^T\ep_N\eta \,. 
\eeq
This bilinear form has the symmetry $\eta\to M\eta$, where
$M^T\ep_N M=\ep_N$. 
This is the defining relation for $M$ to be an element of the 
group $Sp(2N)$. 
  
We will need the Lie algebra of $Sp(2N)$.  Let $M=e^X$, in which
case the above relation implies $X^T\ep_N = -\ep_N X$.  
A linearly independent basis for $X$ is then
$\{1\otimes A, \sigma_x\otimes t_x, \sigma_y\otimes t_y, 
\sigma_z\otimes t_z\}$, where $A$ is an $N\times N$ anti-symmetric
matrix, $\sigma_i$ are the Pauli matrices, and
$t_i$ are $N\times N$ symmetric matrices \cite{Georgi}. 
For any $N$ the algebra $Sp(2N)$ has an 
$SO(3)$ sub-algebra generated by $\sigma_i\otimes 1_N$, which
can in principle be identified with spin.  It also has an 
$SO(N)$ sub-algebra generated by the 
matrices $A$, and an $SU(N)$ sub-algebra generated by
$1\otimes A$ and $\sigma_z\otimes t_z$, where $t_z$ is traceless. 
The Lie algebra $Sp(2)$ is equivalent to 
$SO(3)\cong SU(2)$. 
Note that the $N=2$ case therefore has two different $SO(3)$ 
sub-algebras that could potentially be identified with spin.  

\subsection{Canonical quantization and pseudo-hermiticity}

For simplicity, let us  specialize to the free theory with $N=1$ component 
(the generalization to $N\ne 1$ is straightforward).   
Then the action takes the form
\beq\label{sp5}
   S = \int d^d\xvec\,dt
   \( \d^\mu\eta_1 \d_\mu\eta_2 - m^2 \eta_1\eta_2 \) . 
\eeq
The canonical momenta are $\pi_1=\d_t\eta_2 $ and
$\pi_2=-\d_t\eta_1$, which leads to the  
equal-time anti-commutation relations
\beq\label{sp7}
   \{ \eta_1(\xvec,t), \d_t\eta_2(\xvec',t) \} 
   = - \{ \eta_2(\xvec,t), \d_t\eta_1(\xvec',t) \} 
   = i\delta^{(d)}(\xvec-\xvec') \,.
\eeq

The pseudo-hermiticity of the hamiltonian exactly parallels the
discussion in section II.  If one expands the fields as
\barray\label{sp8}
   \eta_1(x) 
   &=& \int\frac{d^d\kvec }{(2\pi)^{d/2}\sqrt{2\omega_\kvec}}
    \( a_{\kvec+}^\dagger\,e^{-ik\cdot x}
     + a_{\kvec-}\,e^{i k\cdot x} \) , \nonumber\\ 
   \eta_2(x) 
   &=& \int\frac{d^d\kvec}{(2\pi)^{d/2}\sqrt{2\omega_\kvec}}
    \( - a_{\kvec-}^\dagger\,e^{-ik\cdot x}
     + a_{\kvec+}\,e^{i k\cdot x} \) ,
\earray
then eq.~(\ref{sp7}) implies
\beq\label{sp9}
   \{ a_{\kvec-}, a_{\kvec'-}^\dagger \} 
   = \{ a_{\kvec+}, a_{\kvec'+}^\dagger \} 
   = \delta^{(d)} (\kvec -\kvec' ) 
\eeq
In terms of these modes,  the hamiltonian is
\beq\label{sp10}
   H = \int d^d\kvec\,\omega_\kvec
   \( a_{\kvec+}^\dagger a_{\kvec+} - a_{\kvec-} a_{\kvec-}^\dagger 
   \) .
\eeq
The relation between $\eta_2$ and $\eta_1$ is 
$\eta_2=C\eta_1^\dagger C$, 
where $C$ flips the sign of $a_-$ and $a_-^\dagger$, i.e.\
$C a_\pm C=\pm a_\pm$ and $C a^\dagger_\pm C=\pm a^\dagger_\pm$. 
The hamiltonian is pseudo-hermitian, $H^\dagger=C H C$,
in both the free and interacting theory. Thus, for the reasons 
explained in section II, it defines a unitary time evolution.      

\subsection{$Sp(2N)$ conserved charges} 

For $N=1$ the $Sp(2)$ symmetry is 
\beq\label{sp11}
   \eta = \( \matrix{\eta_1 \cr \eta_2 \cr } \) \to e^X \eta \,,
\eeq
where $X=\vec{\alpha}\cdot\vec{\sigma}$ with arbitrary parameters 
$\alpha_i$. The conserved currents following from Noether's construction read
\beq\label{sp12}
   J^z_\mu = - \frac{i}{2} \( \eta_1 \d_\mu\eta_2 
    + \eta_2 \d_\mu\eta_1 \) , \qquad 
   J^+_\mu = -i\eta_1 \d_\mu\eta_1 \,, \qquad
   J^-_\mu = i\eta_2 \d_\mu\eta_2 \,. 
\eeq  
Using the equations of motion and fermion statistics ($\eta_i^2=0$) one readily verifies that $\d^\mu\vec{J}_\mu=0$. 
The conserved charges are then defined as usual as
$\vec{S}=\int d^dx\,\vec{J}_t$.   
In terms of the creation and annihilation operators, 
they take the form
\beq\label{sp13}
   S_z = \inv{2} \int\!d^d\kvec\,\big( a_{\kvec+}^\dagger a_{\kvec+} 
    - a_{\kvec-}^\dagger a_{\kvec-} \big) \,, \qquad
   S^+ = \int\!d^d\kvec\,a_{\kvec+}^\dagger a_{\kvec-} \,, \qquad
   S^- = \int\!d^d\kvec\,a_{\kvec-}^\dagger a_{\kvec+} \,.
\eeq
As expected, these charges satisfy the $Sp(2)\cong SO(3)$ algebra
$[S_z,S^\pm ]=\pm S^\pm$, $[S^+,S^-]=2S_z$ with the identification 
$S^\pm=S_x\pm iS_y$.  

Using the pseudo-hermiticity of the hamiltonian, 
one finds that the above conserved charges 
have the pseudo-hermiticity properties
$(S_z)^\daggerc=S_z$, $(S^\pm)^\daggerc=-S^\mp$.
This implies that pseudo-hermitian conjugation of the $Sp(2)$ generators is an inner-automorphism of the algebra:
\beq\label{pseudoS2}
   \big( \vec{S} \big)^\daggerc 
   = e^{i\pi S_z}\,\vec{S}\,e^{-i\pi S_z} \,.
\eeq

\subsection{Identifying the spin} 

The spin-statistics connection requires that particles with 
half-integer spin under rotations be quantized as fermions.  
Since the Lie algebra $Sp(2N)$ has an $SO(3)$ sub-algebra
generated by $\vec{\sigma}\otimes 1_N$, it is natural to 
try and identify this $SO(3)$ sub-algebra with 
spin (i.e., spacial rotations).   

Again for simplicity let us consider $N=1$ component symplectic fermions.  The one-particle states 
$|\kvec,\pm\rangle = a^\dagger_{\kvec\pm} |0\rangle$ 
of energy $\omega_\kvec$ have spin 
$S_z\,|\kvec,\pm\rangle=\pm\inv{2}\,|\kvec,\pm\rangle$,
so that $a^\dagger_+$ and $a^\dagger_-$
create spin-up and spin-down particles, respectively.  
A further check of this identification comes from considerations of 
time-reversal symmetry,  to which we now turn. 

\subsection{Time reversal and parity}

Let $\CT$ denote the time-reversal operator.  Since $\CT$ is
anti-linear, it can be written as $\CT=UK$, where 
$U$ is unitary and $K$ complex conjugates $c$-numbers:
$Kz=z^* K$.   Consider spin-$\frac12$ particles, where
spin is represented by the Pauli matrices $\vec{\sigma}$.   
Since spin is odd under time reversal,
$\CT\vec{\sigma}\,\CT^{-1} = -\vec{\sigma}$, which implies
$U\vec{\sigma}^* U^\dagger = -\vec{\sigma}$.  
The well-known solution to this equation is $U=\sigma_y$.   
Since time reversal also flips the sign of momentum, 
we are led to define 
\beq\label{time1}
   \CT  a_{\kvec,\pm }  \CT^{-1} 
   = \pm i \, a_{-\kvec,\mp } \,, \qquad
   \CT a^\dagger_{\kvec,\pm } \CT^{-1} 
   = \mp i \, a^\dagger_{-\kvec,\mp} \,.
\eeq
As is well known, due to the anti-unitarity, $\CT^2 = -1$ on 
one-particle states of spin $\frac12$.  

Using the above transformations in eq. (\ref{sp13}), one sees
that the $Sp(2)$ generators have the correct transformation properties
to be identified as rotational spin:
\beq\label{time2}
   \CT \vec{S}\,\CT^{-1} = -\vec{S} \,. 
\eeq
From the form of eq.~(\ref{sp10}) it follows that the hamiltonian
is invariant under time-reversal, i.e.\
$\CT H\,\CT^{-1} = H$.

On the modes,  parity simply flips the sign of momenta, i.e.\
$\CP a_{\kvec,\pm} \CP=a_{-\kvec,\pm}$, 
and similarly for $a^\dagger_\pm$.   The hamiltonian is thus also 
invariant under parity.

\section{Free energy and finite size effects}

Finite-size effects are another probe of the unitarity of a theory.   Let us therefore consider our model embedded
in the geometry of $d$-dimensional flat space with periodic time described by a circle of circumference  $\beta$, i.e., 
$R^d\otimes S^1$.  
We will use the language of quantum statistical mechanics
and identify $\beta = 1/T$ with $T$ being the temperature.   
The $d$-dimensional volume of $R^d$ will be denoted as $V$. 

In Euclidean space the action is 
\beq\label{RG.1}
   S_\chi = \int d^Dx \left[ 
   \d\chibar \d\chi + m^2 \chibar\chi 
   + 4\pi^2 g\,(\chibar\chi)^2 \right] ,
\eeq
where $D=d+1$ is the Euclidean space-time dimension, 
$\d^2=\sum_{i=1}^D \d_i^2$, and as before $\chi$ is
an $N$-component complex fermion field.   
In order to make certain arguments in the sequel,  let us
introduce an auxiliary field $u(x)$ and consider the action
\beq\label{free.1}
   S_{\chi, u} = \int d^Dx \left[ 
   \d\chibar\d\chi + (m^2 + 2\pi\sqrt{g}\,u)\,\chibar\chi
   - \frac{u^2}{4} \right] ,
\eeq
from which the original action $S_\chi$ is recovered when the field 
$u$ is eliminated using its equations of motion.  Since $\chi$ now 
appears quadratically, one can perform the functional integral over
it to obtain
\beq\label{free.2}
   Z = \int{\cal D}\chibar {\cal D}\chi {\cal D}u\,e^{-S_{\chi,u}} 
   \equiv \int{\cal D}u\,e^{-S_{\rm eff}} , 
\eeq
where 
\beq\label{free.3}
   S_{\rm eff} = - N\,\Tr\log[-\d^2 + m^2 + 2\pi\sqrt{g}\,u(x)]
   - \int d^Dx\,\frac{u^2}{4} \,,
\eeq
and we have used the identity $\log\det A=\Tr\log A$.  
Note that if $\chi$ were taken to be a complex bosonic field, then 
the functional
integral would give  $1/\det A$ rather than $\det A$, which amounts
to the replacement $N\to-N$ in $S_{\rm eff}$.   This suggests that some physical quantities in the symplectic
fermion model can be obtaining by flipping the sign of $N$ in its 
bosonic counterpart.   However, we now demonstrate that this
not correct for all
physical quantities; in particular, such a replacement does not hold for the free energy when one takes
into account the proper boundary conditions.  

For the remainder of this section we will consider the 
non-interacting theory ($g=0$). In order to
be able to perform the integrals and to compare with some known
results, we also set the mass $m$ to zero.  
In the free theory the field
$u$ decouples, and the functional integral over $u$ just 
changes the overall normalization of the partition function $Z$.  
Discarding this overall factor one obtains
$Z  = \exp\[ N\,\Tr\log(-\d^2) \]$. 
The free energy density $\CF=-T\log Z/V$ is then simply
$\CF = -\frac{N\,T}{V}\,\Tr\log(-\d^2)$. 
With the Euclidean time compactified on a 
circle of circumference $\beta$, the time component of the momentum 
is quantized, 
$k_0=(2\pi\nu/\beta)$, where $\nu$ is a Matsubara frequency.  The functional trace is then
\beq\label{free.6}
   \Tr\log(-\d^2) = V \sum_\nu \int\intk\,
   \log\[ \kvec^2 + (2\pi\nu/\beta)^2 \] . 
\eeq

In the Euclidean functional integral approach to finite temperature, 
one is required to impose periodic boundary conditions for bosons
and anti-periodic boundary conditions for fermions.   In order
to illustrate an important point,  let us first consider $\chi$ to
be periodic, i.e.\ $\nu$ an integer.   It is a well-known identity
that
\beq\label{free.7}
   \sum_{\nu\in Z} \log\[ \kvec^2 + (2\pi\nu/\beta)^2 \]
   = \beta\omega_\kvec + 2\log(1-e^{-\beta\omega_\kvec}) \,,
\eeq
where $\omega_\kvec=\sqrt{\kvec^2}$ for $m=0$.   
The first term above gives a temperature-independent contribution 
to the free energy, which we can discard by 
defining $\CF$ such that it vanishes at $T=0$. 
The result is
\beq\label{free.8}
   \CF = - 2N T \int\intk\log(1-e^{-\beta\omega_\kvec}) \,.
\eeq
In analogy with black-body formulas in four dimensions, 
let us define a coefficient $c_D$ through
\beq\label{free.9}
   \CF =  -c_D\,\frac{\Gamma(D/2)\,\zeta(D)}{\pi^{D/2}}\,T^D \,, 
\eeq
where $\zeta$ is Riemann's zeta function.  The above normalization
is such that $c_D=1$ for a single free massless boson.  
Performing the integral in
eq.~(\ref{free.8}) one obtains
\beq\label{free.10} 
   c_D = - 2N \qquad \mbox{(periodic b.c.)}
\eeq
in any dimension $D$. 
The negative value of $c_D$ is normally a sign of non-unitarity.   
In two dimensions,
for unitary theories with zero ground-state energy, $c_2$ is the 
Virasoro central charge $c_{\rm vir}$  of the
conformal field theory \cite{BPZ},  and $c_{\rm vir} = -2$ 
is the usual result for a single symplectic fermion 
\cite{symplecticCFT}.
(For a precise, general relation between $c_2$ and $c_{\rm vir}$ see the
end of this section.) 
Note also that $c_D = -2N$ is simply the $N\to -N$ result 
for $N$ free complex massless bosons.  This is to be expected, since
we have computed it using the periodic boundary conditions 
appropriate to bosons.  

The result (\ref{free.10}) is incompatible with the spectrum
of particles computed in section~II.  Clearly this is due to having
taken the wrong boundary conditions for the fields.  For 
anti-periodic boundary conditions,  $\nu$ is half-integer, 
and one has
\beq\label{free.11} 
   \sum_{\nu\in Z+1/2} \log\[ \kvec^2 + (2\pi\nu/\beta)^2 \] 
   = \beta\omega_\kvec + 2\log(1+e^{-\beta\omega_\kvec})
\eeq
instead of (\ref{free.7}), leading to
\beq\label{free.12}
   \CF = -2NT \int\intk\log(1+e^{-\beta\omega_\kvec}) \,. 
\eeq
It is clear from the above expression and basic results in
quantum statistical mechanics that this result corresponds to
$2N$ free fermionic particles with one-particle energies 
$\omega_\kvec$, consistent with the quantization in section~II.   
Performing the integral one finds
\beq\label{free.13}
   c_D = 2N \( 1-\inv{2^{D-1}} \) \qquad 
   \mbox{(anti-periodic b.c.)}
\eeq
in $D$ dimensions.
The thermal central charge $c_D$ is now positive and consistent
with a unitary theory; in fact, it is the same as for $2N$ 
real Dirac fermions.    

A few additional remarks clarifying the  well-studied $2D$ case are 
in order.  Consider the first-order action
\beq\label{fms.1}
   S_{b/c} = \int d^2 x \( b\,\d_\zbar c + \bar b\,\d_z\bar c \) , 
\eeq
where $z$ and $\zbar$ are Euclidean light-cone coordinates.   
Let us define 
  Lorentz ``spin'' with respect to Euclidean rotations,
such that for a holomorphic field $\psi_s(z)$ of spin $s$
\beq\label{fms.1b}
   \psi_s(e^{2i\pi} z) = e^{2i\pi s}\,\psi_s(z) \,.
\eeq  
With this convention, usual Dirac fermions have spin 
$s=\pm\frac12$.  
Note that parameterizing $s=\vartheta/(2\pi)$ implies that 
the spin $s$ is defined modulo $\vartheta=2\pi$.  
Let us assign the following spins to the $b$ and $c$ fields:
\beq\label{fms.1c}
   {\rm spin}(b,c) = \( s+\frac12, -s+\frac12 \) , \qquad
   {\rm spin}(\bar b,\bar c) = \( s-\frac12, -s-\frac12 \) .
\eeq
Then the Virasoro central charge is
known to be $c_2=1-12s^2$ \cite{FMS}. Identifying
\beq\label{fms.2d}
   b = \d_z\chi^\dagger \,, \qquad
   \d_\zbar c = \d_\zbar\chi \,, \qquad
   \bar b = \d_\zbar\chi^\dagger \,, \qquad
   \d_z\bar c = \d_z\chi \,,  
\eeq
then the above first-order action is equivalent to 
our symplectic fermion action with $\chibar \to \chi^\dagger$. 
The above identifications are consistent with 
${\rm spin}(\chi) = -{\rm spin}(\chi^\dagger) = \frac12-s$ 
(with $\vartheta$ defined modulo $2\pi$ as above).  
The usual correspondence
between symplectic fermions and first-order actions
is based on letting $\chi$, $\chi^\dagger$ have spin 0,
which implies $s=\frac12$ and $c_2=-2$ \cite{symplecticCFT}.
However another choice is 
${\rm spin}(\chi) = -{\rm spin}(\chi^\dagger) = \frac12$, 
which gives $s=0$ and $c_2=1$, as we found above in the $2D$ case.   

Another check in two dimensions goes as follows.    The thermal central charge $c_2$ in this section is known to be related to 
the Virasoro central charge $c_{\rm vir}$ by the formula 
$c_{\rm eff}=c_{\rm vir}-24\Delta_{\rm min}$, where 
$\Delta_{\rm min}$ is the minimal conformal scaling dimension.    
In the  twisted (Ramond) sector of the symplectic fermion, the
ground state is known to correspond to   
the twist field with  dimension $\Delta_{\rm min}=-1/8$ 
\cite{Saleur} .  
Since twist fields modify boundary conditions from periodic 
to anti-periodic,  a consistency check is that the value  $c_2 =1$ in eq.~(\ref{free.13}) for $N=1$ should correspond to 
$c_{\rm eff}$ with $c_{\rm vir}=-2$ and $\Delta_{\rm min}=-1/8$, 
and indeed it does.

\section{Renormalization group and critical exponents}

We study the interacting critical point of the
symplectic fermion theory described by the Euclidean action in 
eq.~(\ref{RG.1}), using a position-space approach based on the 
operator product expansion (OPE). In the following section we will
present an alternative derivation of the critical exponents (extended 
to two-loop order) using a technique based on Feynman diagrams 
familiar from quantum field theories for elementary particles.

Consider a general Euclidean action of the form
\beq\label{ope.1}
   S = S_0 + 4\pi^2 g \int d^Dx\,\CO(x) \,, 
\eeq
where $S_0$ is conformally invariant, $g$ is a coupling,
and $\CO$ a perturbing operator.   For our model,   
$S_0$ is the massless free action and $\CO=(\chibar\chi)^2$.
To streamline the discussion, let $\dim{X}$ denote the scaling
dimension of $X$ in energy units, including the non-anomalous 
classical contribution which depends on $D$.  
An action $S$ necessarily
has $\dim{S}=0$.  Using $\dim{d^D\xvec}=-D$, the
classical dimensions of the fundamental couplings and fields are
determined to be 
$\dim{\chi}=(D-2)/2$, $\dim{m}=1$, and $\dim{g}=4-D\equiv\vep$.  
Let us therefore  define the quantum corrections to the 
scaling dimensions $\gamma_\chi$ and $\gamma_m$ as 
\footnote{This convention for $\gamma_m$ differs by a minus sign 
from the one adopted in \cite{spinon}.}
\beq\label{ope.2}
   \dim{\chi}\equiv \frac{D-2}{2} + \gamma_\chi \,, \qquad
   \dim{m}\equiv 1 +  \gamma_m \,.
\eeq
At the critical point, 
$\gamma_\chi$ determines the two-point function 
of the $\chi$ fields via
\beq\label{scale.5}
   \langle\chidag(\xvec)\,\chi(0)\rangle 
   \sim \inv{\absx^{D-2+2\gammachi}} \,.
\eeq
The anomalous dimension $\gamma_m$ can be used
to define a correlation-length exponent $\nu$.   
At the critical point, 
the correlation length diverges as $m\to 0$, i.e.\
$\xi\sim m^{-2\nu}$. Using the fact that $\dim{\xi}=-1$, one has
$-2\nu=\dim{\xi}/\dim{m}$, which implies 
\beq\label{nugammam}
   \nu = \inv{2(1+\gammam)} \,.
\eeq

The lowest-order contributions to the $\beta$-function
and the critical exponents are easily calculated in 
position space. 
At first order in the $\vep$-expansion,  the OPE coefficients
can be computed in four dimensions.  
Consider first the $\beta$-function.   Since $\CO$ is a marginal operator
in $D=4$ (classically $\dim{\CO} = 4$), the OPE gives
\beq\label{ope.3}
   \CO(x)\,\CO(y) = \frac{C}{4\pi^4 |x-y|^4}\,\CO(y) + \dots
\eeq
for some coefficient $C$.  Consider now the correlation function
$\langle X\rangle$ for arbitrary $X$ to second order in $g$: 
\beq\label{ope.4} 
   \langle X\rangle = \langle X\rangle_0
   - 4\pi^2 g \int d^4x\,\langle X\,\CO (x)\rangle_0 
   + \inv{2}\,(4\pi^2 g)^2 \int d^4x\int d^4y\,
   \langle X\,\CO (x)\,\CO(y) \rangle_0 + \dots \,, 
\eeq
where the subscript ${}_0$ indicates that the correlation function 
is computed with respect to the free action $S_0$.   
Using the OPE (\ref{ope.3}) in the above expression
along with $\int_a d^4x/x^4=-2\pi^2\log a$, where $a$ 
is an ultraviolet cut-off, one finds
\beq\label{ope.5}
   \langle X\rangle 
   = \langle X\rangle_0 
   - 4\pi^2 (g+C g^2\log a) \int d^4x\,\langle X\,\CO(x)\rangle
   + \dots \,. 
\eeq
The ultraviolet divergence is removed by letting 
$g\to g(a)=g-C g^2\log a$.   This leads to 
\beq\label{ope.6}
   \beta(g) = - \frac{dg}{d\log a} = - \vep g + C g^2 + \dots \,,
\eeq
where the leading term comes from the classical dimension of $g$.
(Our convention for the sign of the beta-function is as in 
high-energy physics,
where increasing $a$ corresponds to a flow toward low  energy.)

The above calculation easily generalizes to actions of the form
\beq\label{manyg}
   S = S_0 + 4\pi^2 \sum_a g_a \int d^Dx\,\CO_a(x) \,, 
\eeq
which typically arise in anisotropic versions of our model.  
If the perturbing operators satisfy the OPE
\beq\label{manyg.2}
   \CO_a(x)\,\CO_b(y)
   = \sum_c \frac{C^{ab}_c}{4\pi^4 |x-y|^4}\,\CO_c(y) + \dots \,,
\eeq
then the corresponding $\beta$-functions are
\beq\label{manyg.3}
   \beta_a(g) = - \vep g_a + \sum_{b,c} C^{bc}_a\,g_b g_c + \dots \,.
\eeq

Let us return now to our model.  Using the OPE results 
\beq\label{ope.7}
   \chibar_i(x)\,\chi_j(0) 
   =  - \chi_i(x)\,\chibar_j(0)
   \sim - \frac{\delta_{ij}}{4\pi^2 |x|^2}
\eeq
valid in four dimensions to evaluate (\ref{ope.3}) for 
$\CO=(\chibar\chi)^2$, one finds $C= 4-N$ which leads to 
\beq\label{ope.8}
   \beta(g) = - \vep g + (4-N) g^2 + \dots \,. 
\eeq
The model thus has a low-energy fixed point at 
$g_*\approx\vep/(4-N)$.

Consider next the anomalous dimension of the symplectic fermion 
fields and of composite operators built out of these fields.  
Let $\Phi(x)$ denote a field having the following OPE with
the perturbation:
\beq\label{ope.9}
   \CO(y)\,\Phi(x) = \frac{B}{8\pi^4 |x-y|^4}\,\Phi(y) + \dots
\eeq
for some coefficient $B$.   Then to first order in $g$
\barray\label{ope.10}
   \langle\Phi(0)\rangle 
   &=& \langle\Phi(0)\rangle_0 
    - 4\pi^2 g \int d^4y\,\langle\CO(y)\,\Phi(0)\rangle_0  
    + \dots \nonumber\\
   &=& \( 1 + Bg\log a \) \langle\Phi(0)\rangle_0 + \dots
    \approx a^{Bg}\,\langle\Phi(0)\rangle_0 \,. 
\earray
This implies an anomalous contribution $\gamma_\Phi$ to 
$\dim{\Phi}$ given by $\gamma_\Phi=Bg+\dots$.
For the operator $\Phi=\chibar\chi$, the OPE result (\ref{ope.7}) 
implies $B=1-N$.  Finally, using $2\dim{m}+\dim{\chibar\chi}=D$, 
one has
\beq\label{ope.11}
   \gammam = -\inv{2}\,\gamma_{\chibar\chi} 
   = \frac{N-1}{2}\,g + \dots \,.
\eeq

In the sequel it will also be of interest to consider
other fermion bilinears of the form
\beq\label{ope.12}
   n_\tau(x) = \chibar\,\tau\,\chi \,, 
\eeq
where $\tau$ is a traceless matrix. This operator
does not mix with $\chibar\chi$ in the OPE (\ref{ope.9})
and has an independent anomalous dimension $\gamma_n$.
Repeating the above computation,  one finds that because of the 
tracelessness of $\tau$ there is no contribution to this order
proportional to $N$, i.e., $B=1$,  and this leads to 
\beq\label{ope.13}
   \gamma_n = g + \dots \,.
\eeq

To the order we have computed so far,  our results for the $\beta$-function and the anomalous dimension $\gamma_m$ are the same as for
the $O(M)$ Wilson-Fisher fixed point with the substitution $M=-2N$.   This evidently follows from the auxiliary-field construction
in section~IV: bosons versus fermions differ by the overall sign of 
the logarithm of the determinant, which amounts to $N\to -N$ in 
the effective action $S_{\rm eff}$ in eq.~(\ref{free.3}).  
The factor of 2 in $M=-2N$ comes from
the fact that $\chi$ is a complex field, whereas the $O(M)$ model is formulated in terms of $M$ real fields.     
Though one may worry that this equivalence with
exponents of $O(-2N)$ may be spoiled at higher orders for certain operators
whose correlation functions cannot be computed from $S_{\rm eff}$,  
we verify in the next section that the equivalence persists to two-loop order. We thus expect it to hold to all orders in 
perturbation theory \footnote{In \cite{AndreIsing} it was observed
that the known $O(M)$ exponents agree surprisingly well with 
the $N=-M$ (rather than $N=-M/2$) 
symplectic-fermion exponents at lowest order if one identifies 
the anomalous dimension of the $M$-vector order parameter 
$\nvec$ as $\gamma_n=2\gamma_\chi$.  Unfortunately,
the next-order corrections computed in the section~VI spoil this
agreement.}.

\section{Two-loop results}

The simple position-space method of the last section does not extend straightforwardly to higher orders. Here we describe an alternative calculation of the 
$\beta$-function and the anomalous dimensions using Feynman graphs.

We consider the action (\ref{sp1}) of an $N$-component symplectic fermion $\chi$ in Minkowski space and express it in terms of bare parameters $m_0$ and $g_0$ and unrenormalized fields $\chi_0^\alpha$:
\begin{equation}
   S_\chi = \int\!d^Dx \left[
   \partial_\mu\bar\chi_0\,\partial^\mu\chi_0 
   - m_0^2\,\bar\chi_0\chi_0 
   - 4\pi^2 g_0 \big( \bar\chi_0\chi_0 \big)^2 \right] .
\end{equation}
The momentum-space Feynman rule for the four-fermion vertex with incoming fermions $\chi^\alpha$, $\chi^\beta$ and outgoing fermions $\bar\chi^\alpha$, $\bar\chi^\beta$ is $(-8\pi^2 ig_0)$ if $\alpha\ne\beta$, while it vanishes for $\alpha=\beta$ due to the anti-commuting nature of the fields. The momentum-space propagator for the fermion $\chi^\alpha$ is diagonal in component indices
 and given by the ordinary Feynman propagator for a scalar field, $i/(p^2-m_0^2+i0)$. The mass term is kept in our calculations as an infrared regulator. 

\begin{figure}
\begin{center}
\includegraphics[width=0.25\textwidth]{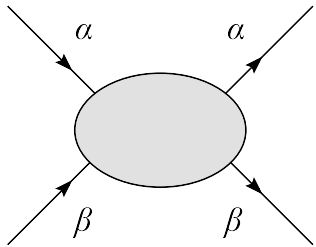}
\end{center}
\vspace{-0.6cm}
\caption{\label{fig:vertex}
Four-fermion vertex function relevant to the calculation of the $\beta$-function.}
\end{figure}

The mass dimensions of the field and coupling are $\dim{\chi_0}=(D-2)/2$ and $\dim{g_0}=4-D$. We work in dimensional regularization and define the renormalized coupling $g(\mu)$ through $g_0=\bar\mu^{4-D} Z_g(\mu)\,g(\mu)$, where $\bar\mu^2\equiv\mu^2\,e^{\gamma_E}/(4\pi)$. The scale $\mu$ acts as an ultraviolet regulator in momentum space. Our renormalization factors will be defined using the modified minimal subtraction ($\overline{\rm MS}$) scheme in $D=4-\vep$ space-time dimensions. 

\subsection{Calculation of the $\beta$-function}

Standard field-theory arguments can be used to show that the $\beta$-function of our model is given by
\begin{equation}\label{beta}
   \beta(g,D) = \frac{dg(\mu)}{d\ln\mu} = \beta(g) + (D-4)\,g \,,
\end{equation}
where
\begin{equation}\label{betadef}
   \beta(g) = g^2\,\frac{dZ_g^{(1)}}{dg}
\end{equation}
is independent of $D$. The quantity $Z_g^{(1)}$ denotes the coefficient of the $1/\vep$ pole in the Laurent expansion of the renormalization factor near $\vep=0$. Throughout, we denote $g\equiv g(\mu)$ unless otherwise noted.

\begin{figure}
\begin{center}
\includegraphics[width=0.17\textwidth]{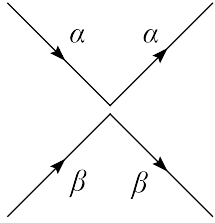}
\end{center}
\caption{\label{fig:vertextree}
Tree-level contribution to the vertex function.}
\end{figure}

\begin{figure}
\begin{center}
\includegraphics[width=0.9\textwidth]{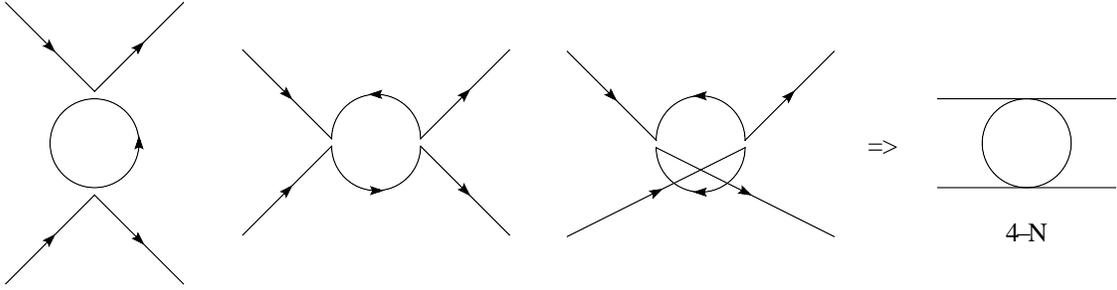}
\end{center}
\vspace{-0.6cm}
\caption{\label{fig:vertex1loop}
One-loop contributions to the vertex function. The three graphs give rise to the same loop topology, shown on the right. The group-theory factors of the individual diagrams are $-(N-2)$, 1, 1, where the minus sign of the first graph results from the closed fermion loop.}
\end{figure}

\begin{figure}
\begin{center}
\includegraphics[width=0.8\textwidth]{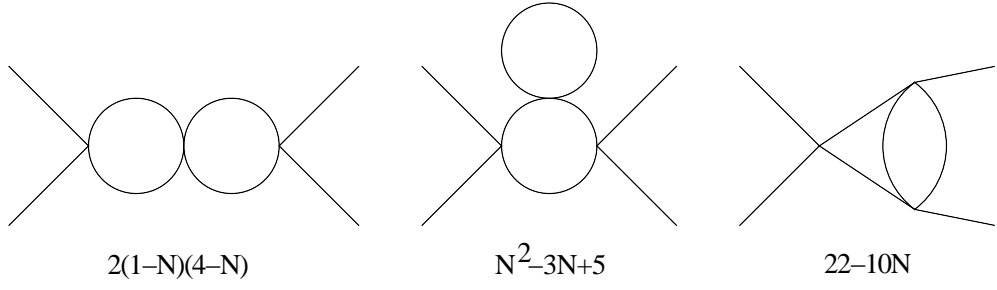}
\end{center}
\vspace{-0.6cm}
\caption{\label{fig:vertex2loop}
Two-loop topologies for the vertex function. Each topology receives contributions from several diagrams. The combined group-theory factors are listed below each graph.}
\end{figure}

The $\beta$-function is obtained from the four-fermion vertex function with ingoing and outgoing component indices $\alpha\ne\beta$, shown in Figure~\ref{fig:vertex}. For simplicity we set the external momenta to zero. The tree-level contribution to the vertex function is given by the elementary vertex shown in Figure~\ref{fig:vertextree}. At one-loop order there is a single loop topology but three different contractions of indices, depicted in Figure~\ref{fig:vertex1loop}, which yield a multiplicity factor of $(4-N)$. At two-loop order there are three loop topologies, whose multiplicities can be obtained by analyzing the various possible contractions. The relevant diagrams and their group-theory factors are depicted in Figure~\ref{fig:vertex2loop}. The two-loop scalar integrals required for our calculation can be obtained from \cite{Caffo:1998du}. 

Adding up the results for the various diagrams we obtain the bare vertex function. We then multiply this result by $Z_\chi^2$ to account for wave-function renormalization, and substitute $m_0^2=Z_{m^2}\,m^2$ and $g_0=\mu^{4-D} Z_g\,g$ to implement mass and coupling-constant renormalization. The renormalization factors $Z_\chi$ and $Z_{m^2}$ are determined from the calculation of the fermion self-energy in the next subsection. By requiring that the renormalized vertex function be finite, we extract
\begin{equation}\label{Zg}
   Z_g = 1 + \frac{4-N}{\vep}\,g + \left[ \frac{(4-N)^2}{\vep^2}
   + \frac{3(3N-7)}{4\vep} \right] g^2 + O(g^3) \,.
\end{equation}
From eq.~(\ref{betadef}) we then obtain for the $\beta$-function
\begin{equation}\label{betares}
   \beta(g) = (4-N)\,g^2 + \frac{3(3N-7)}{2}\,g^3 + O(g^4) \,.
\end{equation}

The $D$-dimensional $\beta$-function in (\ref{beta}) has a non-trivial fixed point at positive coupling given by the solution to the equation
$\beta(g_*)/g_*=4-D=\vep$. At second order in the $\vep$-expansion, we find
\begin{equation}\label{fixed}
   g_* = \frac{\vep}{4-N} + \frac{3(7-3N)}{2(4-N)^3}\,\vep^2 
   + O(\vep^3) \,.
\end{equation}

\subsection{Calculation of the self-energy}

Next we need the anomalous dimension of the fermion mass and field. They follow from a two-loop calculation of the self-energy $\Sigma(p^2,m_0^2)$ in the vicinity of the mass shell ($p^2=m^2$). Here $m_0$ and $m$ are the bare and renormalized mass parameters, respectively. The relevant relations are
\begin{equation}
   m^2 = m_0^2 + \Sigma(m^2,m_0^2) \,, \qquad
   Z_\chi^{-1} = 1 - \Sigma'(m^2,m_0^2) \,,
\end{equation}
where the prime denotes a derivative with respect to the first argument, $p^2$. Inserting here $m_0^2=Z_{m^2} m^2$, one finds for the renormalization factors 
\begin{equation}
   Z_{m^2} = 1 - \frac{\Sigma(m^2,Z_{m^2}m^2)}{m^2} \,, \qquad
   Z_\chi^{-1} = 1 - \Sigma'(m^2,Z_{m^2}m^2) \,.
\end{equation}

\begin{figure}
\begin{center}
\includegraphics[width=0.7\textwidth]{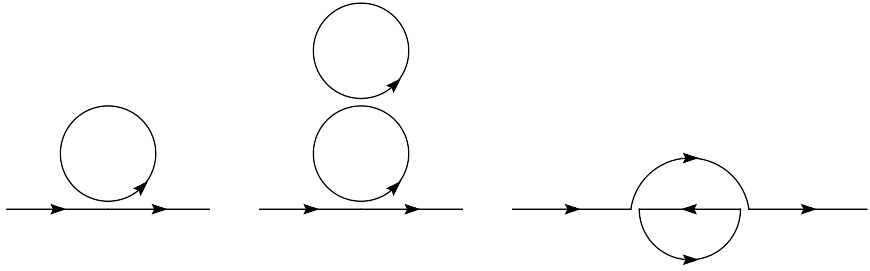}
\end{center}
\vspace{-0.6cm}
\caption{\label{fig:self}
One- and two-loop diagrams contributing to the fermion self-energy.}
\end{figure}

At one-loop order there is a single tadpole graph to evaluate, while at two-loop order we have a double tadpole diagram and the sunrise diagram, see Figure~\ref{fig:self}. After accounting for coupling-constant renormalization using eq.~(\ref{Zg}), we obtain
\begin{eqnarray}
   Z_{m^2} &=& 1 + \frac{1-N}{\vep}\,g
    + (1-N) \left( \frac{5-2N}{2\vep^2}
    - \frac{5}{8\vep} \right) g^2 + O(g^3) \,, \nonumber\\
   Z_\chi &=& 1 - \frac{1-N}{8\vep}\,g^2 + O(g^3) \,. 
\end{eqnarray}
The anomalous dimensions of the mass and field are given by
\begin{equation}
   \gamma_{m^2} = - g\,\frac{dZ_{m^2}^{(1)}}{dg} \,, \qquad 
   \gamma_\chi = - \frac{g}{2}\,\frac{dZ_\chi^{(1)}}{dg} \,,
\end{equation}
where in the second relation we take into account that $\chi_0=\sqrt{Z_\chi}\,\chi$. We find
\begin{eqnarray}\label{gammares}
   \gamma_{m^2} = 2\gamma_m &=& (N-1)\,g\,\bigg( 1 - \frac54\,g \bigg) 
    + O(g^3) \,, \nonumber\\
   \gamma_\chi &=& \frac{(1-N)}{8}\,g^2 + O(g^3) \,.
\end{eqnarray}
The anomalous dimension of the field starts at two-loop order. Instead of the anomalous dimension of $m^2$ one could compute the anomalous dimension of the fermion bilinear $\bar\chi\chi$, which is given by $\gamma_{\bar\chi\chi}=-\gamma_{m^2}$. 

Evaluating our expressions at the fixed-point value of the coupling yields
\begin{eqnarray}
\label{values}
   \gamma_m
   &=& \frac{(N-1)}{2(4-N)}\,\vep \left[
    1 + \frac{22-13N}{4(4-N)^2}\,\vep \right] + O(\vep^3) \,, \nonumber\\ 
   \gamma_\chi &=& \frac{(1-N)}{8(4-N)^2}\,\vep^2 + O(\vep^3) \,.
\end{eqnarray}

As a crosscheck of our results, we note that the two-loop expressions for the $\beta$-function and anomalous dimensions obtained in eqs.~(\ref{betares}) and (\ref{gammares}) go over to the corresponding results of $O(M)$ scalar field theory (see e.g.\ \cite{Zinn}) if we identify $M=-2N$. Likewise, the fixed-point of the fermionic $\beta$-function in eq.~(\ref{fixed}) is related to the Wilson-Fisher fixed point \cite{WilsonFisher} by the same replacement rule. In some sense, our symplectic fermion theory may thus be considered as an analytic continuation of scalar $O(M)$ field theory to negative $M$. However, as emphasized in section~IV, this simple correspondence
is not expected to hold for all physical quantities.  

\subsection{Renormalization of the composite bilinear $n_\tau$}

The diagrams contributing to the renormalization of the composite operator $n_\tau=\bar\chi\,\tau\,\chi$ defined in eq.~(\ref{ope.12}) can be obtained by inserting this operator into the one- and two-loop graphs for the fermion self-energy shown in Figure~\ref{fig:self}. In the evaluation of these graphs it is important to use that the matrix $\tau$ is traceless. Writing the bare current as $(n_\tau)_0=Z_n n_\tau$, we obtain
\begin{equation}
   Z_n = 1 - \frac{g}{\vep}
   + \left( - \frac{3-N}{2\vep^2} + \frac{5-N}{8\vep} \right) g^2 
   + O(g^3) \,.
\end{equation}
The anomalous dimension of the current is thus
\begin{equation}
\label{comp}
   \gamma_n = g - \frac{5-N}{4}\,g^2 + O(g^3) \,.
\end{equation}
At the fixed point, this yields
\begin{equation}
\label{gammanvecep}
   \gamma_n = \frac{\vep}{4-N} + \frac{(2-N)(11+N)}{4(4-N)^3}\,\vep^2
   + O(\vep^3) \,.
\end{equation}
This result will become important for the discussion in the following section.

\section{Possible applications}

In this section,  we speculate on possible applications of
the above results.    The most interesting context is 
quantum criticality in $d=2$ spacial dimensions (i.e., 
$D=3$).  In the broadest terms, since the particles
have spin $\inv{2}$, our model can describe a quantum critical theory
of spinons.   

Whereas the Mermin-Wagner theorem rules out 
continuous phase transitions at finite temperature in $d=2$,
zero temperature quantum phase transitions continue to be
of great interest.   The best studied example is the 
quantum phase transition in $2d$ Heisenberg magnets,
which is in the universality class of the $O(3)$ Wilson-Fisher
fixed point \cite{Halperin,Ye}.  
One feature of our model is that it can describe
quantum critical points wherein the magnetic order parameter
$\nvec$ is a composite operator in terms of the more fundamental
fermion fields.   This could in principle have applications to
quantum phase transitions in the anti-ferromagnetic phase of
Hubbard-like models, where the magnetic order parameter is 
a bilinear in the electron fields.   As stated in the introduction,
if such electrons were described by the Dirac theory,  the
four-fermion interactions are irrelevant and do not generally
lead to a low-energy interacting fixed point.   
That this is different in the case of symplectic fermions was the primary motivation for our work.  

Let us first review the definitions of the exponents 
for the usual Wilson-Fisher fixed point.   The order parameter
is an $M$-component real vector $\nvec$ with action
\beq\label{SWF}
   S_{\rm WF} = \int d^Dx \[ \d\nvec\cdot\d\nvec
    + m^2\,\nvec\cdot\nvec + \lambda\,(\nvec\cdot\nvec)^2 \] . 
\eeq  
Some of the exponents are defined with respect to perturbations
away from the critical point.
Namely,  consider  
\beq\label{sdeform}
   S = S_* +  \int d^Dx \( t\,\CO_\vep + \vec{B}\cdot\nvec \) ,
\eeq
where $S_*$ is the critical theory,  $\CO_\vep$ is the 
``energy operator'',  and $\vec{B}$ a magnetic field.  
For applications to classical statistical mechanics in $3D$ one
identifies $\CO_\vep=\nvec\cdot\nvec$ and $t=T-T_c$, so that
$t\propto m^2$.     
The usual definition of the correlation-length exponent 
$\nu_\epsilon$ via $\xi\sim t^{-\nu_\vep}$ then leads to
\beq\label{dqc.4} 
   \nu_\vep = \inv{\dim{t}} = \inv{D-\dim{\CO_\vep}} \,.
\eeq

The second fundamental exponent is related to the 
scaling dimension of $\nvec$.  It is conventional
to parameterize this with $\eta$ in the form 
\beq\label{dqc.8}
   \dim{\nvec} = \frac{D-2}{2} + \frac{\eta}{2} \,, 
\eeq
so that the two-point function at the critical point 
scales as 
\beq\label{dqc.8b}
   \langle\nvec(x)\cdot\nvec(0)\rangle 
   \sim \inv{|x|^{D-2+\eta}} \,.
\eeq
The convention for the leading contribution  to
$\dim{\nvec}$ comes from the action (\ref{SWF}), 
which implies that classically $\nvec$ has dimension $(D-2)/2$.
The parameter $\eta$ is then given as $\eta=2\gamma_\nvec$, where
$\gamma_\nvec$ is the quantum anomalous correction to
the scaling dimension of $\nvec$.   
 
The third exponent characterize the one-point function of
$\nvec$ via
\beq\label{dqc.5}
   \langle\nvec\rangle\sim t^\beta ~ \sim B^{1/\delta} ,
\eeq
which leads to 
\beq\label{dqc.6} 
   \beta = \frac{\dim{\nvec}}{\dim{t}} 
   = \frac{\nu_\vep}{2}\,(D-2+\eta) \,.
\eeq
Treating $\vec{B}$ as a coupling gives $\dim{B}+\dim{\nvec}=D$, from which it follows that
\beq\label{dqc.7} 
   \delta = \frac{\dim{B}}{\dim{\nvec}}  
   = \frac{D+2-\eta}{D-2+\eta} \,.
\eeq

For the remainder of this section we consider the special case 
where $N=2$.  
In discussions of deconfined quantum criticality \cite{Senthil}, 
for the case of $O(3)$ symmetry, the 3-vector $\nvec$ is represented
as a bilinear  in ``spinon fields'' $\chi$, i.e.\
$\nvec=\chi^\dagger\vec{\sigma}\,\chi$,
where here $\chi$ is an $N=2$ component complex field 
and $\vec{\sigma}$ are the Pauli matrices. Note that $\vec n$ is 
an example of an operator $n_\tau$ defined in (\ref{ope.12}). 
For $N=2$, 
the above representation of $\nvec$ is consistent irrespective
of whether $\chi$ is a boson or fermion,  so we will treat
both cases in parallel. 
If $\nvec$ is a rotational 3-vector,  then $\chi$  is
a spin-$\inv{2}$ doublet.  This identification of spin is different
than in section III, and is instead based on the $SU(2)$
sub-algebra generated by $\{1\otimes A, \sigma_z\otimes t_z\}$,
where $t_z$ is traceless and symmetric (see section III for notations). 

Simple considerations based on the renormalization group point  
to a possibly special role played by the $N=2$ fermion theory. 
Suppose a model formulated in terms of an $O(3)$ $\nvec$ field
is asymptotically free in the ultraviolet region.  
Then the coefficient of the free energy $c_D$ described in 
section~IV equals $3$ for any $D$.   
On the other hand, free $\chi$ fields give $c_3 = 3N/2$ by
formula (\ref{free.13}) in $3D$. which is the same as for the $O(3)$
$\nvec$ theory when $N=2$.   Therefore, the free energies match up
properly in the ultraviolet for an $O(3)$ $\nvec$ field and
an $N=2$ symplectic fermion.  

The model proposed in \cite{Senthil} is based on the $CP^1$ representation
of the non-linear $O(3)$ sigma model,  which involves 
an auxiliary $U(1)$ gauge field.  The $CP^1$ model is then
modified by relaxing the non-linear constraint 
$\nvec\cdot\nvec=\chi^\dagger\chi=1$ and making the 
gauge field dynamical by adding a Maxwell term, effectively
turning the model into an abelian Higgs model.    
One appealing feature of
this model is that because of the equivalence of $CP^1$ and $O(3)$ 
non-linear sigma models (at least classically), without the Maxwell 
term one has an explicit map between the non-linear $\nvec$ field  
and $\chi$-field actions.  Though this model is a 
natural candidate for a deconfined quantum critical point,  
unfortunately the fixed point is difficult to study perturbatively, 
so it has not  been
possible to accurately compare exponents with the simulations
reported in \cite{Motrunich,Sandvik}.   

Let us broaden the notion of deconfined quantum criticality 
to simply refer to a quantum critical point  for an $O(3)$
vector order-parameter $\nvec$, where $\nvec$ is composite
in terms of spinon fields $\chi$.   
The fields $\chi$ are interpreted as the fundamental 
underlying degrees of freedom, and the critical theory 
$S_*$ in eq. (\ref{sdeform}) is the critical theory for $\chi$.
The critical exponents $\nu$, $\eta$, $\beta$, and $\delta$ 
defined above
are then related to scaling dimensions of composite operators in
the $S_*$ theory.    

The anomalous dimension of $\nvec$ in the epsilon expansion 
follows from the results in section~VI and will be denoted
$\gamma_\nvec$ in what follows. 
Let us first consider the case where $\chi$ is a bosonic field. 
As explained above, the exponents for bosonic verses fermionic
theories are simply related by $N \to -N$.   
Specializing eq.~(\ref{gammanvecep}) to $N=-2$ in $3D$ ($\vep =1$) 
one finds $\gamma_\nvec\approx 0.21$ for bosonic $\chi$.
This is quite large compared to the analogous result 
$\gamma_\nvec \approx 0.02$ for
the $O(3)$ Wilson-Fisher fixed point. 
(The latter is well-known;  as explained in section V,  it
corresponds to $\gamma_\chi$ at $N=-3/2$.)  
For $\chi$ a fermion, interestingly the $O(\vep^2)$ correction
to the leading one-loop result vanishes for $N=2$,  so that 
$\gamma_\nvec=1/2$.  We can give an alternative estimate of 
$\gamma_\nvec$ by simply substituting $g_*=11/16$ from 
eq.~(\ref{fixed}) into (\ref{comp}) without expanding in $\vep$. 
This gives $\gamma_\nvec\approx 0.33$ for fermionic $\chi$.
For both the bosonic and fermionic cases, the largeness of
$\gamma_\nvec$ compared with the usual $O(3)$ fixed point
is due to the compositeness of $\nvec$.  
If we naturally identify $\CO_\vep$ with $\chibar\chi$, 
then the correlation-length exponent $\nu_\vep$ is given by
eqs.~(\ref{nugammam}) and (\ref{values}) evaluated with $N=\pm 2$. 
This leads to
\beq\label{nuvalues}
   \nu_\vep\approx 0.75 ~~ ({\rm {bosonic}}) \,, \qquad
   \nu_\vep\approx 0.42 ~~ ({\rm fermionic}) \,.
\eeq

There  are at least two  difficulties encountered if we attempt to
compare with existing numerical simulations, such as those in 
\cite{Motrunich,Sandvik}.  
The main one is that the simulations are performed with
an action for the $\nvec$ field or for local lattice
spin variables $\vec{S}_i$ with a Heisenberg-like hamiltonian,
rather than with fundamental spinon degrees of freedom $\chi$,
and we do not have a direct map between the two descriptions.  
In particular, in a theory with fundamental $\chi$ fields
and the compositeness relation 
$\nvec=\chi^\dagger\vec{\sigma}\,\chi$, since $\chi$ has classical dimension $(D-2)/2$, one has
\beq\label{dqc.20}
   \dim{\nvec} = D-2 + \gamma_\nvec \,.
\eeq
Comparing with eq.~(\ref{dqc.8}) one finds $\eta=D-2+2\gamma_\nvec$
rather than the usual relation $\eta=2\gamma_\nvec$.  In other 
words, $\eta$ now contains
a purely classical contribution of $D-2$, and this was
used to argue that the $\eta$ exponent was large in \cite{Senthil}.    On the other hand, simulations 
based on $\nvec$-field actions effectively force the
classical contribution to $\dim{\nvec}$ to be $(D-2)/2$
as in the Wilson-Fisher theory,  suggesting that 
simulations measure $\eta = 2 \gamma_\nvec$.   
In the fermionic theory,  support for this idea
comes from the fact that the two lowest orders of the
$\vep$ expansion give $\gamma_\nvec = 1/2$ in $D=3$. 

The  other difficulty  is that, unlike for temperature
phase transitions where $t=T-T_c$,  in the context of
zero temperature quantum critical points it is not obvious what plays
the role of the parameter 
$t$, or equivalently, the energy operator $\CO_\vep$ that determines
the correlation-length exponent.  In this context, $t$ is 
a parameter in the hamiltonian that is tuned to the critical point. 
The most natural choice
is $\CO_\vep=\chibar\chi$ which implies $t = m^2$ and leads
to eq.~(\ref{nuvalues}).  However, another possibility could be 
$t=m$, which would lead to twice the values in eq.~(\ref{nuvalues}),
corresponding to $\nu_\vep=1/(1+\gammam)$. 

The above difficulties prevent us from establishing a 
definite connection with the simulations in \cite{Motrunich,Sandvik}.
In fact, the exponents for deconfined quantum criticality 
are currently controversial, since the two above works disagree 
strongly on the value of the exponent $\eta$.   
However, we point out that if one identifies
$\eta=2\gamma_\nvec$, then our computed 
exponents are not inconsistent with some of the exponents in
\cite{Motrunich,Sandvik}, although
the comparison is not conclusive. 
More specifically, the work \cite{Motrunich}
reports $\eta\approx 0.6$--0.7 and $\nu=0.8$--1.0. 
On the other hand, 
for a different model \cite{Sandvik},  it was found
that $\eta=0.26\pm 0.03$ and $\nu=0.78\pm 0.03$.
Both simulations are consistent with $\nu_\vep$ 
in eq.~(\ref{nuvalues}) for a {\it bosonic} spinon
and $t=m^2$.  However, they are also consistent with
a fermionic spinon with $t= m$.        
Our formulas give
$\eta\approx 0.67$ and $\eta\approx 0.42$ for fermions and 
bosons, respectively.

\section{Conclusions}

To summarize,  we proposed that spin-$\inv{2}$ particles can
be described by a symplectic fermion quantum field theory 
as an alternative to the Dirac theory if one demands only rotational 
invariance rather than the full Lorentz invariance.   
The resulting lagrangian has a form resembling that of a scalar field
but with the ``wrong'' statistics.  A hidden $Sp(2N)$ symmetry
allows identification of spin via an $SO(3)$ subgroup for any $N$,
so that the statistics of the field
is consistent with the spin-statistics connection for spin-$\inv{2}$ 
particles.
The hamiltonian is pseudo-hermitian, and this is sufficient
to guarantee a unitary time evolution.   
The usual spin-statistics theorem for this kind of field
theory is circumvented, because the proof of the latter does not
allow for a pseudo-hermitian hamiltonian.  

We have analyzed the renormalization-group properties and critical exponents of the symplectic fermion model up to two-loop order. 
The anomalous dimensions and $\beta$-function of the $Sp(2N)$ model
are related to those of the $O(M)$ Wilson-Fisher model by setting $M=-2N$. 
This correspondence between $O(M)$ and $Sp(2N)$ models 
does not hold for all physical properties however.  
In addition to the usual exponents,
we have also computed exponents for
fields that are bilinear in the fundamental spinon fields.

The potentially most interesting possible applications of our theory 
are to quantum critical spinons in $d=2$ spacial dimensions.  
We have computed 
the critical exponents for ``magnetic'' order parameters 
that are quadratic in the spinon fields, as in models of
deconfined quantum criticality.   Comparison with 
existing numerical simulations is to some extent favorable,  
but not yet conclusive.

\subsection*{Acknowledgments}

One of us (A.L.) would like to thank C. Henley, A. Ludwig,
F. Nogeira, N. Read, 
S. Sachdev, A. Sandvik, T. Senthil, and J. Sethna for useful discussions. 
This research was supported in part by the National Science 
Foundation under Grant PHY-0355005.

\end{document}